\documentclass[aps,prb,preprint,showpacs]{revtex4}
\usepackage{CJK}
\usepackage{graphicx}% Include figure files
\usepackage{dcolumn}% Align table columns on decimal point
\usepackage{bm}
\usepackage[utf8]{inputenc}
\usepackage[english]{babel}
\usepackage{graphicx}
\graphicspath{{images/}{../images/}{../../images}}
\usepackage{braket}
\usepackage{blindtext}
\usepackage{chemformula}
\usepackage{amsmath}
\usepackage{amssymb}
\usepackage{subfiles}
\usepackage[autostyle=true,autopunct=true]{csquotes}
\usepackage{mathtools} % for the sign \coloneqq -> :=
\usepackage{cleveref}
\usepackage{siunitx}

\newcommand*{\deriv}[2]{\dfrac{d#1}{d#2}}
\newcommand*{\magm}[1]{\mu_{#1}}
\newcommand*{\vmag}[1]{\mathbf{m}_{#1}}

\newcommand*{\gyro}[1]{\gamma_{#1}}
\newcommand*{\vecm}[1]{\mathbf{m}_{#1}}
\newcommand*{\normm}[1]{m_{#1}}

\newcommand*{\vechmfa}[1]{\mathbf{H^{MFA}_{#1}}}

\newcommand*{\relml}[1]{\Gamma_{#1,\parallel}}
\newcommand*{\relmt}[1]{\Gamma_{#1,\perp}}

\setcitestyle{numbers,square}
\begin{document}
\title{Landau-Lifshitz-Bloch equation for ferrimagnets with higher-order interaction}

\author{Marco Menarini}
\email{menarini.marco@gmail.com}
\author{Vitaliy Lomakin}%
\affiliation{%
 Department of Electrical and Computer Engineering, Center for Memory and Recording Research, University of California, San Diego, La Jolla, California 92093}%
%\affiliation{Department of Physics, University of California, San Diego, La Jolla, California 92093-0319}%

\date{\today}% It is always \today, today,

\begin{abstract}
We present a micromagnetic formulation for modeling the magnetization dynamics and thermal equilibrium in ferrimagnetic materials at low and elevated temperatures. The formulation is based on a mean field approximation (MFA). In this formulation, the ferrimagnet is described micromagnetically by two coupled sublattices with corresponding interactions, including inter- and intra-sublattice micromagnetic exchange as well as four-spin interactions described as an inter-sublattice molecular field with a cubic dependence of the magnetization. The MFA is used to derive a Landau Lifshitz Bloch type equation for ferrimagnetic material, including cases with a ferromagnetic - antiferromagnetic phase transitions. For validation, the results obtained via the presented model are compared with recent experimental data for phase transitions in FeRh.
\end{abstract}
\maketitle

%\tableofcontents

\section{\label{sec:introduction}Introduction}
There is an  increased interest in using antiferromagnetic (AF) materials for creating reliable and compact sources of coherent signals in the THz frequency. This is enabled due to the fact that the frequency of antiferromagnetic resonances  $\omega_{AFMR}$ can reach the THz range, significantly exceeding the frequency of ferromagnetic resonances\cite{gomonay2014spintronics,baltz2018antiferromagnetic}. Several devices for spin torque oscillators have been proposed that leverage the strong inter-sublattice AF exchange as the source of the THz signal \cite{khymyn2017antiferromagnetic} and using the spin current to induce a canting angle between the two sublattices. Such devices have been proposed as possible THz frequency comb-generators to be used as artificial neurons for neuromorphic computing due to their fast response time and threshold behaviour \cite{khymyn2018ultra}. 

Recently, Medapalli et al. \cite{medapalli2020femtosecond} showed that it is possible to optically generate a THz pulse in a FeRh/Pt bi-layer. In the experiment, an ultrafast laser pulse excites metamagnetic FeRh injecting a spin-current into the non-magnetic Pt interface that is, then, converted into a spin-current via the inverse spin Hall effect \cite{nakayama2012geometry,cheng2016terahertz}. The spin current in the AF state can originate from a precessional response of FeRh during a partial phase transition induced by the laser \cite{menarini2019micromagnetic}. Such transformation occurs on a sub picosend time scale, much faster than any lattice expansion \cite{ju2004ultrafast}. The phase transition occurs due to the competition between bilinear and the Rh mediated biquadratic exchange interactions in an effective spin Hamiltonian \cite{mryasov2005magnetic}. Bilinear and biquadratic exchange energies strongly depend on the temperature. Using atomistic simulations, it is possible to reproduce such phase transitions by including both the bilinear and biquadratic exchanges \cite{barker2015higher}. 

However, despite the computing power of modern computers, to model realistic structures, a coarse-grained model for the dynamic of the magnetization is desirable. The Landau-Lifshitz-Bloch (LLB) equation of motion for macroscopic magnetization vectors \cite{Garanin1997} has been used to accurately model the behaviour of complex magnetic structures at high temperatures. Its usability has been extended by Atxitia et al. \cite{Atxitia2012} to ferrimagnets with two sublattices. However, this model cannot describe phase transition between ferromagnetic and antiferromagnetic states as observed in experiments \cite{ju2004ultrafast,thiele2003ferh} and may miss additional effects related to the inter-sublattice micromagnetic exchange interactions.

In this paper, we present an LLB formulation for ferrimagnetic materials introducing effects of higher-order exchange and show that they are necessary to model a metamagnetic AF/FM transitions driven by temperature. We derive a macroscopic equation for the magnetization dynamics of two-sublattice metamagnetic systems with higher order exchange valid in the entire temperature range. As a concrete test case, we consider metamagnetic FeRh particles. FeRh is modelled as two sublattices, each with its length and direction, coupled via an inter-sublattice exchange. We use the mean-field approximation (MFA) to derive a macroscopic equation for the magnetization of each sublattice. We study the mean field energy of the system to better understand the phase transition and validate the model against the experimental results.

\section{\label{sec:atommodel} Mean Field Approximation of a two sublattice system with higher-order interactions}
\begin{figure}
    \centering
    \includegraphics[width=\textwidth]{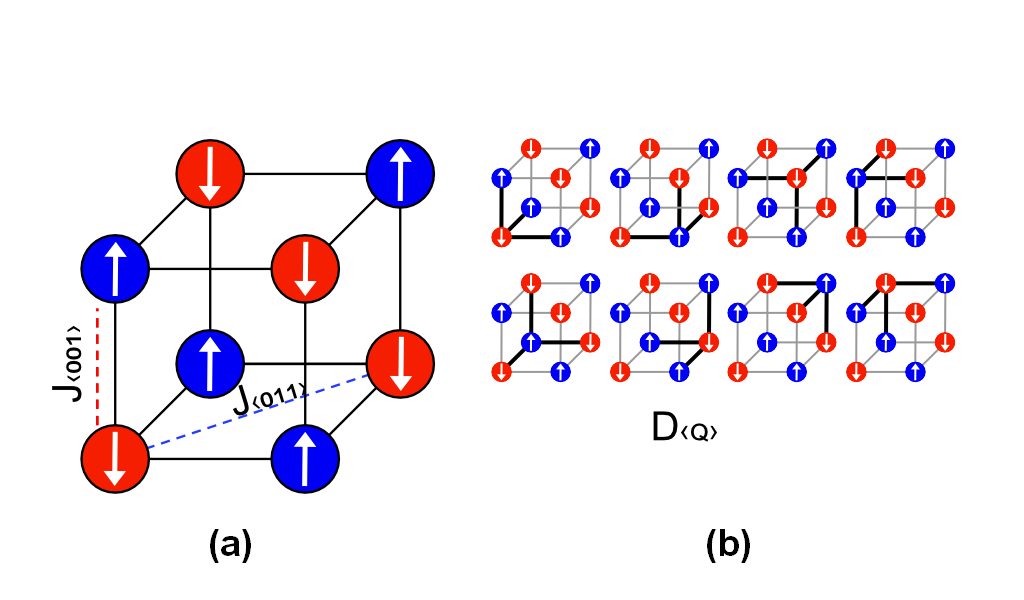}
    \caption{Simplified model of the unit cell (a) with the nearest-neighbor exchange (red dashed line) $J_{\langle001\rangle}$ and the second nearest-neighbor exchange (blue dashed lines) $J_{\langle011\rangle}$. In (b) eight 4-spin cyclical interactions inside the unit cell (thick dark lines) are shown.}
    \label{fig:4spinmodel}
\end{figure}

We start by consider an atomistic model for an FeRh ferrimagnet as used by Barker et al. \cite{barker2015higher}. The effective Hamiltonian $\mathcal{H}$ contains only the degrees of freedom of a simple cubic (sc) Fe lattice , with the effect of the induced Rh moment included into effective Fe-Rh-Fe interactions. The Hamiltonian is augmented by the applied field $\mathbf{H}$ and uniaxial anisotropy:
\begin{equation}
\label{eq:hamiltonian}
\begin{split}
\mathcal{H}&=-\sum_{i}{\mu_i\mathbf{H}\mathbf{S}_i}+\sum_{ij}{J_{ij}\eta_{ij} \left(S_{i,x}S_{j,x}+S_{i,y}S_{j,y}\right)}
-\sum_{ij}{J_{ij}\mathbf{S}_i\mathbf{S}_j}\\
&+\frac{1}{3}\sum_{i,j,k,l}{D_{ijkl}
\left[
\left(\mathbf{S}_i\mathbf{S}_j\right)
\left(\mathbf{S}_k\mathbf{S}_l\right)+
\left(\mathbf{S}_i\mathbf{S}_k\right)
\left(\mathbf{S}_j\mathbf{S}_l\right)+
\left(\mathbf{S}_i\mathbf{S}_l\right)
\left(\mathbf{S}_k\mathbf{S}_j\right)
\right]}
\end{split}\, .
\end{equation}
Here, $\mathbf{S}_i$ is the normalized spin vector of the atoms $i$ and $\mu_i$ is its magnetic moment. $J_{ij}$ are the Heisenberg exchange interactions (bilinear), including the direct Fe-Fe and indirect Fe-Rh-Fe contributions. $D_{ijkl}$ are the four-spin exchange (biquadratic) coefficients, which only have contributions from the Fe-Rh-Fe interactions. The parameter $\eta_{ij}\ll 1$ defines the strength of the anisotropy in the direction perpendicular to the easy axis \cite{Garanin1997}. For the Heisenberg exchange interactions, only the nearest neighbors and the second nearest neighbors inside the unit cell are considered (\cref{fig:4spinmodel}(a)). The cyclical four-spin interaction inside each unit cell is given by pairwise interactions between the 3 nearest neighbors converging on one of the vertices of the sc lattice (\cref{fig:4spinmodel}(b)).

The free energy of the system described by $\mathcal{H}$ in \cref{eq:hamiltonian} can be given as $F=-T\ln\mathcal{Z}$, where $\mathcal{Z}$ is the partition function and $T$ is the temperature. In the mean-field approximation we consider each spin on a site $i$ as an isolated spin subjected to the effective field due to the mean values of the neighboring spins. 

Since in the AF state the nearest neighbors tend to be antiparallel to each other and the second nearest neighbors tend to be parallel and taking into account the symmetry of the system, we can consider this mean field as the field produced by the two sublattices $\boldsymbol{m}_{A,i}=\langle\mathbf{S}_{A,i}\rangle$ and $\boldsymbol{m}_{B,i}=\langle\mathbf{S}_{B,i}\rangle$. The mean-field Hamiltonian is then obtained from \cref{eq:hamiltonian} as:
\begin{equation}
    \label{eq:mfahamiltonian}
    \mathcal{H}^{MFA}=\mathcal{H}_{00}-\sum_{i}\sum_{\mu=A,B}{{\magm{\mu}\mathbf{H}_{\mu,i}^{MFA}\mathbf{S}_{\mu,i}}}\, ,
\end{equation}
The term $\mathcal{H}_{00}$ is given by
\begin{equation}
\begin{split}
    \label{eq:hamiltonian00}
    \mathcal{H}^{00}&=
    \frac{J_{\langle 011\rangle}}{2} 
    \sum_{ij}{\sum_{\mu=A,B}{\left(\mathbf{m}_{\mu,i}\mathbf{m}_{\mu,j}\right)}
    }+
    \frac{J_{\langle 011\rangle}}{2}\sum_{ij}{\sum_{\mu=A,B}{\sum_{k=x,y}{
    \eta_{\mu}
    \left(\mathbf{m}_{\mu,i}\cdot\mathbf{\hat{e}}_{k}\right)\left(\mathbf{m}_{\mu,j}\cdot\mathbf{\hat{e}}_{k}\right)
    }}}\\
    &+
    \frac{J_{\langle 001\rangle}}{2}\sum_{ij}{
    \left(\mathbf{m}_{A,i}\mathbf{m}_{B,j}\right)}
    -12D_{\langle Q\rangle}\sum_{i}{
    \sum_{\substack{\mu=A,B \\ \mu\neq \nu}}{\left(\mathbf{m}_{\mu,i}\mathbf{m}_{\mu,i}\right)
    \left(\mathbf{m}_{\mu,i}\mathbf{m}_{\nu,i}\right)
    }}
\end{split}\, ,
\end{equation}
where $J_{\langle 011\rangle}$ is the inter-sublattice exchange coefficient, $J_{\langle 001\rangle}$ is the intra-sublattice exchange coefficient, and $\mathbf{\hat{e}}_{k}$ is the unit vector in the direction of $k=x,y$. The molecular field for the two sublattices $\mu,\nu=A,B$ is given by
\begin{align}
\begin{split}
    \label{eq:hmfa}
    \magm{\mu}\mathbf{H}_{\mu,i}^{MFA}&=\magm{\mu} \mathbf{H}+
    J_{\langle 011\rangle} 
    \sum_{j}{\mathbf{m}_{\mu,j}}+
    J_{\langle 011\rangle}\sum_{j}{\sum_{k=x,y}{
    \eta_{\mu}\left(\mathbf{m}_{\mu,j}\cdot\mathbf{\hat{e}}_{k}\right)\mathbf{\hat{e}}_{k}}}\\
    &+
    \frac{J_{\langle 001\rangle}}{2}\sum_{j}{
    \left(\mathbf{m}_{\nu,j}\right)}
    -8D_{\langle Q\rangle}
    \left(\mathbf{m}_{\mu,i}\mathbf{m}_{\nu,i}\right)
    \mathbf{m}_{\mu,i}
    -4D_0\left(m_{\mu,i}^2+m_{\nu,i}^2\right)\mathbf{m}_{\nu,i}
\end{split}\, .
\end{align}
 The solution of the one-spin problem in \cref{eq:mfahamiltonian} leads to
\begin{equation}
\label{eq:onespin}
F=\mathcal{H}_{00}-NT\ln(4\pi)-T\sum_{i}{\sum_{\mu}{\Lambda\left(\xi_{\mu,i}\right)}}\, ,\qquad \Lambda\left(\xi\right)=\ln\left(\frac{\sinh\left(\xi\right)}{\xi}\right)\, ,
\end{equation}
where $N$ is the total number of spins, $\xi_{\mu,i}=\left|\boldsymbol{\xi}_{\mu,i}\right|$ is the reduced field for the sublattice $\mu$ and spin $i$ with $\boldsymbol{\xi}_{\mu,i}=\magm{\mu}\beta\mathbf{H}_{i}^{MFA}$, and $\beta=1/T$, where the temperature $T$ is given in the units of energy. The MFA free energy in \cref{eq:onespin} can be minimized with respect to the average magnetization $\mathbf{m}_{\mu,i}$ to find the equilibrium solution of the system. 

If we consider the continuum limit we can go from the sums in \cref{eq:hamiltonian00,eq:hmfa} to volume integrals. For small anisotropy and assuming small changes of the magnetization between spins in the same sublattices, we can rewrite the short-range interaction between the nearest neighbors and second nearest neighbors as:
\begin{align}
\label{eq:exchangeappears2}
&\sum_{j}{J_\mathbf{\langle 001 \rangle}\mathbf{m}_{\nu,j}}\approx J_1\mathbf{m}_{\nu,i}+A_{ex,\mu\nu}\Delta\mathbf{m}_{\nu,i}\, ,\\
\label{eq:exchangeappears1}
&\sum_{j}{J_\mathbf{\langle 011 \rangle}\mathbf{m}_{\mu,j}}\approx J_2\mathbf{m}_{\mu,i}+A_{ex,\mu\mu}\Delta\mathbf{m}_{\mu,i}\, .
\end{align}
Here, $\Delta$ is the Laplace operator acting on the sublattice magnetization $\mathbf{m}_{A}(\mathbf{r})$. In addition, $J_1=zJ_\mathbf{\langle 001 \rangle}$ is the average of the exchange interactions for $z=6$ nearest neighbors in the sc lattice and $J_2=qJ_\mathbf{\langle 011 \rangle}$ is the average over the second nearest neighbors with $q=12$. 
For the sc lattice, the exchange constants are given by $A_{ex,\mu\mu}=2J_2a^2_0/q$ and $A_{ex,\mu\nu}=J_1a^2_0/z$, where $a_0$ is the lattice spacing assumed to be the same in both directions. 

 Substituting \cref{eq:exchangeappears1,eq:exchangeappears2} in \cref{eq:hamiltonian00,eq:hmfa} and taking the continuum limit in \cref{eq:onespin}, one obtains:
\begin{equation}
    \label{eq:finalenergy}
    \begin{split}
    \frac{F}{J_2}&=\frac{1}{v_0}
    \int{
        d\mathbf{r}
        \sum_{\substack{\mu=A,B \\ \mu\neq \nu}}
            {\left\{
                \frac{1-6d\left(\mathbf{m}_\mu\mathbf{m}_{\nu}\right)}{2}
                m_\mu^2
                +j\frac{\mathbf{m}_\mu\mathbf{m}_\nu}{2}
                +\frac{\left(\mathbf{m}_{\mu},\mathbf{h}^{eff}_{\mu}-\mathbf{h}_{\mu}\right)}{2}
                -\frac{1}{\beta J_2}\Lambda(\xi_\mu)
        \right\}}
    }\\
    &-\frac{NT}{J_2}
    \end{split}\, ,
\end{equation}
where $v_0$ is the unit-cell volume, $j=J_1/(2J_2)< 1$ is the normalized inter-sublattice exchange coefficient, and $d=4D_{\langle Q\rangle}/J_2< 1/6$ is the normalized four-spins coefficient. The reduced field and the normalized effective fields for the sublattice $\mu$ are given by
\begin{align}
    \label{eq:redfieldA}
    &\mathbf{\xi_\mu}=\beta J_2
    \left\{
    \left[1-2d\left(\mathbf{m}_{\mu}\mathbf{m}_{\nu}\right)\right]
    \mathbf{m}_{\mu}+
    \left[\frac{j}{2}-d\left(\mathbf{m}_{\mu}^2+\mathbf{m}_{\nu}^2\right)\right]
    \mathbf{m}_{\nu}+\mathbf{h}^{eff}_{\mu}
    \right\}\, ,\\
    \label{eq:redfieldB}
    &\mathbf{h}^{eff}_{\mu}=\mathbf{h}_\mu+\frac{A_{ex,\mu\mu}}{J_2}\Delta\mathbf{m}_{\mu}+\frac{A_{ex,\mu\nu}}{J_2}\Delta\mathbf{m}_{\nu}-\eta_\mu\sum_{k=x,y}{\left(\mathbf{m}_{\mu}\cdot\mathbf{\hat{e}}_{k}\right)\mathbf{\hat{e}}_{k}}\, ,
\end{align}
where $\mathbf{h}_\mu=\mu_\mu\mathbf{H}/J_2\ll 1$ is the normalized applied field and $\mathbf{h}_{eff_{\mu,i}}$ acting on the sublattice $\mu$. 

The reduced field given in \cref{eq:redfieldA} and the effective field given in \cref{eq:redfieldB} can be used to formulate an LLB equation for ferrimagnets with higher order interaction, as we show in the next section.

\section{\label{sec:micromodel} LLB equation for higher order ferrimagnet}
To derive a two component LLB model, we follow the procedure outlined by Atxitia et al.  \cite{Atxitia2012}. By substituting  \cref{eq:exchangeappears1,eq:exchangeappears2} into the \cref{eq:hmfa} we obtain the mean-field approximation of the molecular field:
\begin{subequations}
        \label{eq:hmfa1}
    \begin{align}
        &\mathbf{H}_{\mu,i}^{MFA}=\mathbf{H}^{eff}_{\mu, i}+
        \mathbf{H}^{\parallel}_{E_{\mu, i}}+
        \mathbf{H}^{\perp}_{E_{\mu, i}}
        \, ,\\
        &\magm{\mu}\mathbf{H}^{eff}_{\mu, i}=\magm{\mu}\mathbf{H}
        +A_{ex,\mu\mu}\Delta m_{\mu,i}
        +A_{ex,\mu\mu}\Delta m_{\mu,i}
        -\magm{\mu}H_{K,\mu}\sum_{k=x,y}{\left(\mathbf{m}_{\mu,i}\cdot\mathbf{\hat{e}}_k\right)\mathbf{\hat{e}}_k}\, , \\
        &\mathbf{H}^{\parallel}_{E_{\mu, i}}=
        \frac{J^\parallel_{\mu,i}}{\magm{\mu}} \mathbf{m}_{\mu,i}\, ,\\
        \label{eq:hmfa1_Jperp}
        &\mathbf{H}^{\perp}_{E_{\mu, i}}=-\frac{J^\perp_{\mu,i}}{\magm{\mu}}
        \frac{
        \vmag{\nu,i}\times\left(\vmag{\nu,i}\times\vmag{\mu,i}\right)
        }{m_{\nu,i}^2}\, ,\\%\mathbf{m}_{\nu,i}\left(1-\Theta(m_{\nu,i},m_{\mu,i})\right)\, ,\\
        &J^\parallel_{\mu,i}=J_2\left[\left(1-2d\left(\mathbf{m}_{\mu,i}\mathbf{m}_{\nu,i}\right)\right)+
        \left(\frac{j}{2}-d\left(m_{\mu,i}^2+m_{\nu,i}^2\right)\right)\Theta(m_{\nu,i},m_{\mu,i})\right]\, ,\\
        &J^\perp_{\mu,i}=J_2
        \left[\frac{j}{2}-d\left(m_{\mu,i}^2+m_{\nu,i}^2\right)\right]\, ,
    \end{align}
\end{subequations}
where $d=4D_{\langle Q\rangle}/J_2$, $j=J_1/J_2$, $H_{K,\mu}=J_2\eta_\mu/\magm{\mu}$ is the anisotropy field, and $\mathbf{H}^{\parallel}_{E_{\mu, i}}$ and $\mathbf{H}^{\perp}_{E_{\mu, i}}$ are the intra-sublattice parallel and perpendicular exchange, respectively. Given two vectors $\mathbf{v}_A$ and $\mathbf{v}_B$, the function
\begin{equation}\label{eq:scalingtheta}
    \Theta(v_{A},v_{B})=
    \frac{\mathbf{v}_{A}\cdot \mathbf{v}_{B}}{v^2_{B}} \, , 
\end{equation}
is the projection of the vector $\mathbf{m}_A$ in the direction of the vector $\mathbf{m}_B$. We substitute the MFA for the field in \cref{eq:hmfa1} into the dynamic formulation of the mean magnetization obtained through the Fokker-Planck equation \cite{Garanin1997}. The corresponding set of coupled LLB equations for each sublattice $\mu$ is given by
\begin{equation}
    \label{eq:llb}
    \deriv{\vecm{\mu}}{t}=\gyro{\mu}\left[\vecm{\mu}\times\vechmfa{\mu}\right]
    -\relml{\mu}\left(1-\frac{\vecm{\mu}\vecm{0,\mu}}{\normm{\mu}^2}\right)\vecm{\mu}
    -\relmt{\mu}\frac{\left[\vecm{\mu}\times\left(\vecm{\mu}\times\vecm{0,\mu}\right)\right]}{\normm{\mu}^2}\, ,
\end{equation}
where $\gyro{\mu}$ is the gyromagnetic ratio, $\relml{\mu}$ and $\relmt{\mu}$ are the longitudinal and transverse relaxation rates, and the instantaneous equilibrium magnetization $\vecm{0,\mu}$ is given by
\begin{equation}
    \label{eq:mv0}
    \vecm{0,\mu}=B(\xi_{\mu})\frac{\boldsymbol{\xi}_{\mu}}{\xi_{\mu}}\, ,\qquad \boldsymbol{\xi}_{\mu}=\beta\mu_{\mu}\vechmfa{\mu}\, .
\end{equation}
Here, $\xi_{\mu}=\left|\boldsymbol{\xi}_{\mu}\right|$ is the reduced field and $B\left(x\right)=\coth(x)-1/x$ is the Langevin function. The parallel and perpendicular relaxation rates are given by
\begin{align}
    \label{eq:parrelax}
    \Gamma_{\mu,\parallel}&=\Lambda_\mu
    \frac{B\left(\xi_{\mu}\right)}{\xi_{0,\mu}B^\prime\left(\xi_{\mu}\right)}\, ,\\
    \label{eq:perrelax}
    \Gamma_{\mu,\perp}&=\frac{\Lambda_\mu}{2}
   \left[
   \frac{\xi_{\mu}}{B\left(\xi_{\mu}\right)}-1
   \right]\, ,
\end{align}
where $\Lambda_\mu=2\gamma_\mu\lambda_\mu/\beta\mu_\mu$ is the characteristic diffusion relaxation rate given by the Neel attempt frequency with the atomistic damping constant $\lambda$.

Equation \cref{eq:llb} with \cref{eq:mv0} and \cref{eq:hmfa1} can be directly used for numerical modeling. However, it is possible to rewrite it in a more compact form if the parallel intra-sublattice exchange is large in comparison with the other components of the MFA field (i.e., $\left|H^{\parallel}_{E,\mu}\right|\gg \left|H^{eff}_{\mu}\right|$ and $\max{\left[j,4d-j\right]}\ll 2$), which is valid in the entire range of temperatures for many ferromagnetic and ferrimagnetic materials \cite{Garanin1997}. Using this approximation, we can expand the Langevin equation to the first order of the Taylor series around $H^{\parallel}_{E,\mu}$:
\begin{align}
    \label{eq:expansion}
    &\mathbf{m}_{0,\mu}\approx\frac{B(\xi_{0,\mu})}{m_\mu}\mathbf{m_\mu}+\mu_{0,\mu}\beta B^\prime(\xi_{0,\mu})\frac{\left(\mathbf{m}_{\mu}\mathbf{H}^{eff}_{\mu}\right)\mathbf{m}_{\mu}}{m_\mu^2}
    \, ,
\end{align}
where $\xi_{0,\mu}=\beta J^{\parallel}_{\mu} m_{\mu}$. Using \cref{eq:expansion}, we can  write the LLB equation in the standard form:
\begin{equation}
    \label{eq:stdllb}
    \begin{split}
    \deriv{\vecm{\mu}}{t}&=
    \gamma_{\mu}\left[\mathbf{m}_\mu\times\left(\mathbf{H}^{eff}_{\mu}+\mathbf{H}^{\perp}_{E_{\mu}}\right)\right]
    -\gamma_{\mu}\alpha_{\parallel,\mu}
    \left(\frac{1-B\left(\xi_{0,\mu}\right)/m_\mu}
    {\mu_{0,\mu}\beta B^\prime\left(\xi_{0,\mu}\right)}-
    \frac{\mathbf{m}_\mu\mathbf{H}^{eff}_{\mu}}{m^2_\mu}\right)\mathbf{m}_\mu\\
    &
    -
    \gamma_\mu\alpha_{\perp,\mu}
    \frac{\mathbf{m}_\mu\times\left[\mathbf{m}_\mu\times\left(\mathbf{H}^{eff}_{\mu}+\mathbf{H}^{\perp}_{E_{\mu}}\right)\right]}{m^2_\mu},
    \end{split}
\end{equation}
where the parallel and perpendicular damping coefficients are functions of the temperature and the angle between the two sublattices:
\begin{equation}
\label{eq:aparallel}
\alpha_{\parallel,\mu}=\frac{2\lambda_\mu T}{\beta J_{\parallel}}
\, ,\qquad  
\alpha_{\perp,\mu}=
\lambda_\mu
\left[1-\frac{T}{\beta J_\parallel}\right]
\, .
\end{equation}
Since the perpendicular intra-sublattice exchange in \cref{eq:stdllb} only  appears in the precessional and the longitudinal damping terms, the contribution of $\vmag{\nu,i}$ in the direction of $\vmag{\mu,i}$ in the cross product $\vmag{\mu,i}\times\vmag{\nu,i}$ is zero by geometrical reasoning, and equation \cref{eq:hmfa1_Jperp} can be rewritten using the triple vector product identity and the function $\Theta$  defined in \cref{eq:scalingtheta}:
\begin{equation}
    \mathbf{H}^{\perp}_{E_{\mu, i}}=\frac{J^\perp_{\mu,i}}{\magm{\mu}}
        \Theta(m_{\mu,i},m_{\nu,i})\vmag{\nu,i}\, .
\end{equation}

\section{\label{sec:simul} Results}
In this section, we use the MFA of the energy and LLB formulations developed in \cref{sec:atommodel,sec:micromodel} to study the phase transition in an example material. We choose FeRh for the readily available experimental literature \cite{ju2004ultrafast,thiele2003ferh,kouvel1962anomalous} and atomistic simulations \cite{barker2015higher}. First, we consider the equilibrium conditions by minimizing the free energy with respect to the magnetization to obtain the critical point at which we have the transition between the AF and FM states. Then, we study the magnetization behaviour via the modified LLB equation and compare it with experimentally results.

\subsection{\label{sec:resultenergy} Energy and thermal equilibrium analysis}
To study the equilibrium conditions, we consider an isotropic case with $\mathbf{h}^{eff}_{\mu}=0$. This case allows obtaining an analytical solution for the energy and demonstrating the model use in a clear way, including the AF to FM transitions.
% Added the reason here
An external field or anisotropy can also be added. These additional field components only change the preferential direction of the system and their effects can be studied numerically via the perturbation theory, e.g., as done for the ferromagnetic case in Ref.~\cite{kachkachi2001magnetic}.

\begin{figure}
    \centering
    \includegraphics[width=\textwidth]{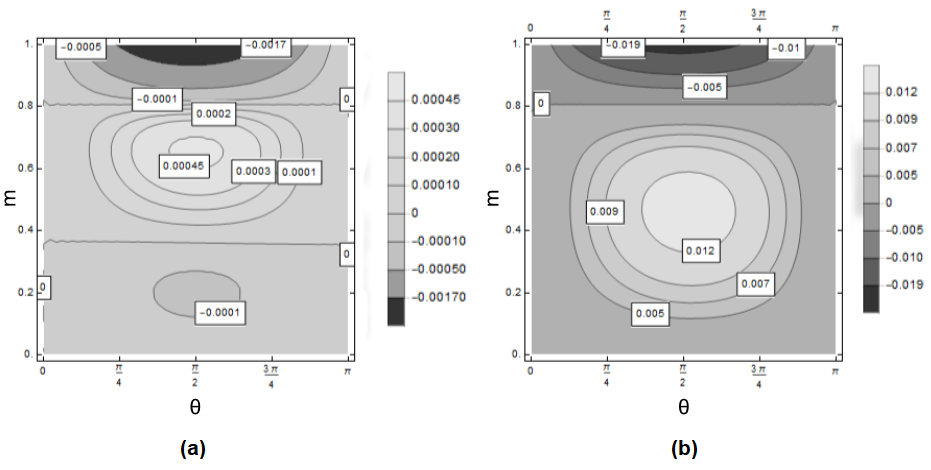}
    \caption{Derivative of the free energy with respect to (a) the magnetization length and (b) the angle between the magnetization of the sublattices as a function of the magnetization length $m$ and the angle $\theta$.}
    \label{fig:energy_contour}
\end{figure}
We minimize the terms between the brackets in \cref{eq:finalenergy} with respect to the magnetization vector. In the absence of an external field the system is symmetric with respect to the polar angle $\phi$. The energy minimization can be accomplished by obtaining the values of $p_{cr}=\{m_{A,cr},m_{B,cr},\theta_{A,cr},\theta_{B,cr}\}$ for which $\left.\partial F/\partial \mathbf{m}_{\mu}\right|_{p=p_{cr}}=\partial F/\partial m_{\mu}\mathbf{\hat{r}}+1/m\cdot \partial F/\partial \theta_\mu \boldsymbol{\hat{\theta}}=0$ and $\left. \partial^2 F/\partial\mathbf{m}_i\partial\mathbf{m}_j\right|_{p=p_{cr}}>0$.
If we use one of the sublattices as the reference of our system, we can set $\theta_\nu=0$ and obtain a solution with respect to the angle only for $\theta_\mu=\theta$, which allows reducing the system with 6 degrees of freedom to an equivalent system with 3 degrees of freedom for the vector $p=\{m_A,m_B,\theta\}$. The first derivative of $\Lambda(x)$ is the Langevin function $B(x)=\text{coth}(x)-1/x$ and the reduced field is given by
\begin{equation}
    \label{eq:redfield1}
    \xi_\mu=\beta J_2\sqrt{
    \begin{matrix}
    \left[m_\mu
    \left(
        1-2d m_\mu m_\nu \cos(\theta_\mu-\theta_\nu)
    \right)\cos(\theta_\mu)
    +
    m_\nu
    \left(
        \frac{j}{2}-2d \left(m^2_\mu +m^2_\nu\right)
    \right)\cos(\theta_\nu)
    \right]^2 +\\
    \left[m_\mu
    \left(
        1-2d m_\mu m_\nu \cos(\theta_\mu-\theta_\nu)
    \right)\sin(\theta_\mu)
    +
    m_\nu
    \left(
        \frac{j}{2}-2d \left(m^2_\mu+m^2_\nu\right)
    \right)\sin(\theta_\nu)
    \right]^2 
    \end{matrix}
    }\, .
\end{equation}
\begin{figure}
    \centering
    \includegraphics[width=0.5\textwidth]{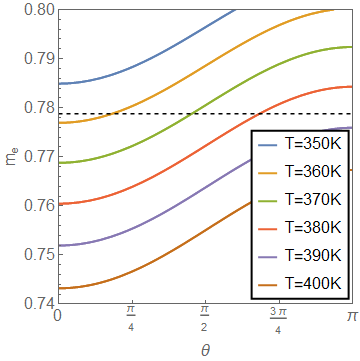}
    \caption{Equilibrium magnetization $m_e$ as a function of the angle $\theta$ for different temperature. The black dashed line is the critical equilibrium magnetization $m_{cr}=\sqrt{j/4d}$}
    \label{fig:funcangle}
\end{figure}
Due to the symmetry of the system, at the equilibrium we expect to have $m_A=m_B=m_e$, and $\xi_A=\xi_B=\xi_e$. This is true when $\theta=n\pi$ with $n=0,1,2,\dots$ or when $j=4dm_e^2$. The minimum condition of the energy \cref{eq:onespin} for $m_e$ and $\xi_e$ leads to a modified Curie-Weiss equation:
\begin{equation}
    \label{eq:energy_contour}
    \mathbf{m}_{e,\mu}=B\left(\xi_e(T,m_e,\theta_\mu,\theta_\nu)\right)\frac{\boldsymbol{\xi}_e(T,m_e,\theta_\mu,\theta_\nu)}{\xi_e(T,m_e,\theta_\mu,\theta_\nu)}\, .
\end{equation}

We define the value of the critical equilibrium magnetization as $m_{cr}=\sqrt{j/4d}$. When the magnetization of the two sublattices is above the critical value $m>m_{cr}$ the effective exchange between the two sublattices is AF, and the equilibrium condition is reached for $\theta=\pi$. When the magnetization of the two sublattices is below the critical value $m<m_{cr}$ the equilibrium is reached for $\theta=0$ and the material is in the FM state (\cref{fig:energy_contour}).

Since the equilibrium magnetization $m_e$ and the the effective exchange are functions of the angle between the two sublattices (\cref{fig:funcangle}), it is possible for the two sublattices to be in either the AF or FM configuration depending on the previous history of the system (i.e. hysteretic behaviour of the phase transition).

\subsection{\label{sec:resultnumeric}LLB analysis}
To validate the LLB model, we first study the phase transition observed in FeRh  as a function of the temperature \cite{thiele2003ferh,kouvel1962anomalous} and, then, the timescale of phase transition as a function of $\lambda$. We conclude this section by presenting an application of our model, for a theoretical material exhibiting a ferrimagnetic to ferromagnetic first order phase transition. A good example of such material are the Heusler alloys, which show similar ferri- to ferromagnetic transition close to room temperature \cite{ovichi2014ferri}. Similarly, to FeRh, the phase transition in these alloy can be explained via the interaction between the bilinear and the biquadratic exchange \cite{simon2020noncollinear,bosu2008biquadratic}. 

We define the magnetization as the mean of the magnetization in the two sublattices $\mathbf{M}=(\mathbf{M}_A+\mathbf{M}_B)/2M_0$ and the Néel vector as $\mathbf{M}_N=(\mathbf{M}_A-\mathbf{M}_B)/2M_0$, where $M_0=(M_{S,A}+M_{S,B})/2$ and $M_{S,A}, M_{S,B}$ are the saturation magnetization in the two sublattices \cite{chiang2019absence}. For sublattices with the same magnetic moments, such as FeRh, the magnetization and Néel vector are defined as:
\begin{equation}
    \mathbf{M}=\frac{\mathbf{m}_A+\mathbf{m}_B}{2}\, ,\quad\mathbf{M}_N=\frac{\mathbf{m}_A-\mathbf{m}_B}{2}\, ,
\end{equation}
where $\mathbf{m_A}$,$\mathbf{m_B}$ are the magnetization vectors of the sublattice A and B, respectively, normalized with respect to the saturation magnetization $M_{S,A}=M_{S,B}=M_{S}$. 

Since by using the MFA, we neglected the higher order wave fluctuations, we update the parameters obtained from the atomistic model for FeRh \cite{barker2015higher} by a correction factor $\epsilon$ to match the experimental results quantitatively. The correction factors are given in Tab.~\ref{tab:dataferh}. To avoid using a correction factor, we can obtain $J_2$, $j$, and $d$ directly from the experimental data for $T_C$ and the phase transition temperatures.

\begin{table}
\begin{tabular}{ |c|c|c|c| } 
\hline
 & Value & $\epsilon$  & Unit   \\
\hline
$J_2$ & $ 2.44035\times 10^{-20}$ &$0.7025$ &  J \\ 

$j$ &$7.7743\times 10^{-22}$  &$1.5202$ & \\ 
$d$ &$3.2046\times 10^{-22}$  &$1.7081$   &\\ 
$m_{cr}$ & $0.7788$& &\\ 
$\mu_{Fe}$   & $3.15$& &$\mu_b$\\ 
\hline
\end{tabular}

\caption{\label{tab:dataferh}Corrected magnetic parameters and correction factor $\epsilon$.}
\end{table}

We first consider an isotropic particle of $5\text{nm}\times5\text{nm}\times5\text{nm}$ initially in the AF state with a critical atomistic constant $\lambda=1$. The temperature is increased step-wise from 1K up to 720K. At every thermal step, the system is let to relax for $40\si{ps}$ to reach the equilibrium. The magnetization length and the antiferromagnetic Néel vector length are obtained by averaging over a $20\si{ps}$ period after the system reaches the equilibrium.

The particle is let to evolve according to the dynamics described in \cref{eq:stdllb} augmented with the uncorrelated thermal field acting on the longitudinal and perpendicular relaxation terms of each sublattices described in Ref.~\cite{menarini2020thermal}. The system is integrated numerically using a semi-implicit scheme \cite{mentink2010stable} to accurately solve the stochastic differential equation in a way that satisfied the Stratonovich calculus \cite{kloeden1992stochastic}.

\begin{figure}
    \centering
    \includegraphics[width=0.9\textwidth]{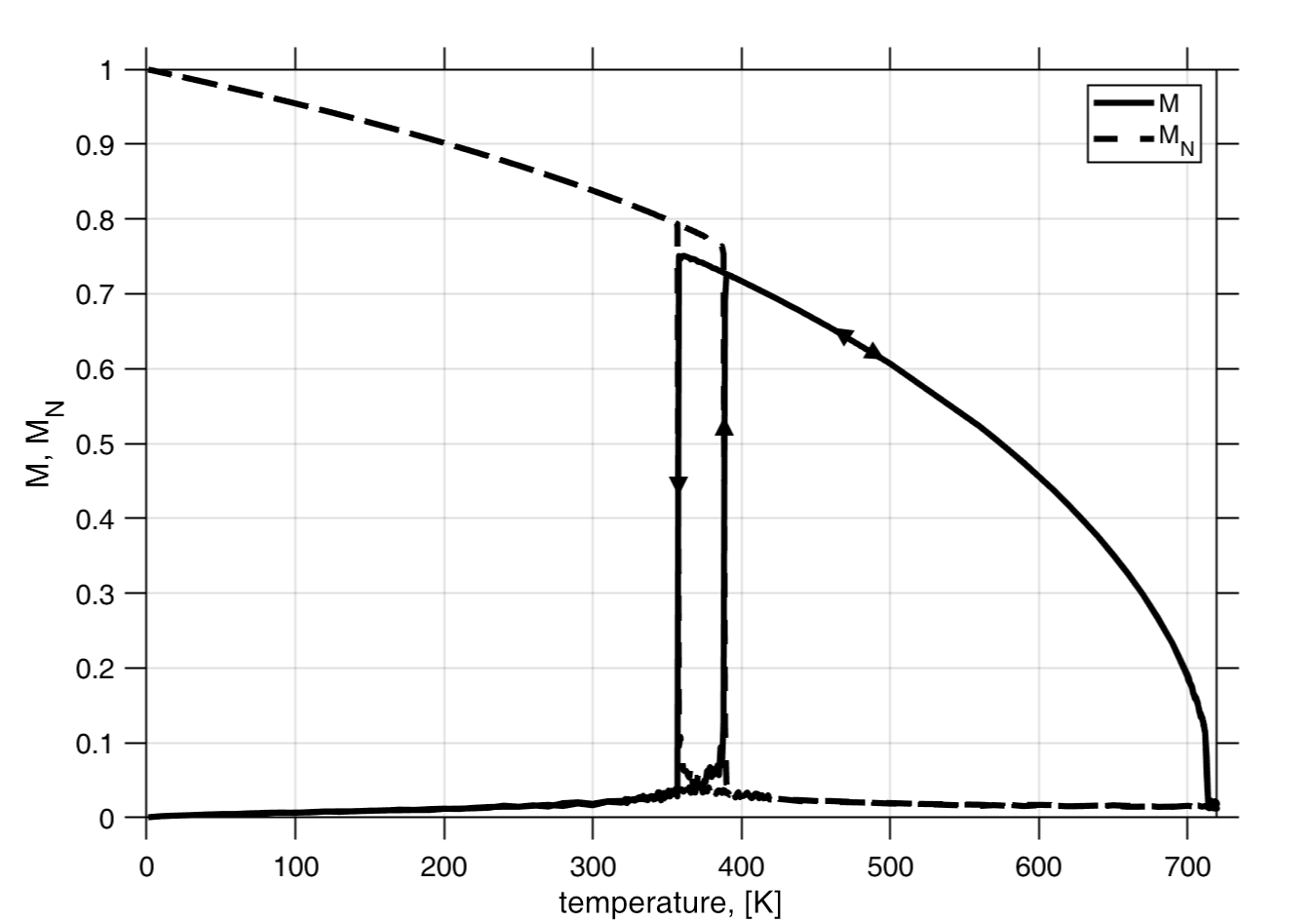}
    \caption{Magnetization (solid line) and Néel vector (dashed line) for an isotropic macrospin of FeRh as a function of the temperature.}
    \label{fig:hysteresis}
\end{figure}

Figure \ref{fig:hysteresis} shows the equilibrium magnetization as a function of the temperature. Similar to what is done with FM materials, we can relate the Curie temperature of the material with the effective exchange constant in each sublattice $J_2\left(1+j\right)=3k_b T_C$ \cite{kachkachi2001magnetic}. As shown in \cref{sec:resultenergy}, the material is susceptible to a phase transition when the magnetization in the two sublattices is close to the critical value $m_{cr}=\sqrt{j/4d}$. The magnetization of the material in the region close to the transition temperature $T_M$ (i.e., $m_{\mu,e}\sim m_{cr}$) is a function of the magnetization history of the material. 
%\begin{figure}
%    \centering
%    \includegraphics[width=\textwidth]{magnetization_vs_state.png}
%    \caption{Equilibrium magnetization in each sublattice as a function of the %temperature for the FM state (solid line) and the AF state (dashed line).}
%    \label{fig:mestate}
%\end{figure}
This hysteretic behaviour, expected from the analysis of the free energy and observed in the experiments \cite{thiele2003ferh,kouvel1962anomalous,barker2015higher} can be explained by looking at the interaction between the reduced parameters $j$ and $d$. Due to the presence of the the four-spin exchange, the equilibrium magnetization in the two phases is a function of the material state %\cref{fig:mestate} 
and it is given by
\begin{equation}
    m_{e,AF/FM}=B(\xi_{AF/FM})\, ,\qquad \xi_{AM/FM}=\beta J_2\left[1\mp\left(\frac{j}{2}-3d m^2\right)\right]m^2\, .
\end{equation}
At lower temperature $T\ll T_M$ the contribution of the four-spin interactions is dominant (i.e., $j\ll 6dm^2$) and $m_{e,AF}>m_{e,FM}$ while at higher temperatures the cubic component of the four-spin interactions drops faster than the linear component of the nearest neighbors (i.e. $j\gg 6dm^2$), leading to $m_{e,AF}<m_{e,FM}$. Depending on the initial phase of the system, the magnetization in the two sublattices reaches the critical point $m_{cr}$ at different temperatures, depending on the initial configuration of the system, hence the hysteresis loop. By controlling the parameters $j$ and $d$, it is possible to engineer the position and the width of the phase transition.

To study the dynamical response of the macro-magnetic particle to a rapid change of temperature, we consider the effect of a sub-picosecond laser pulse modelled as a Gaussian thermal pulse. In FeRh, the initial magnetization response due to an ultrafast thermal pulse is observed in the first $500\si{fs}$, significantly faster than the lattice expansion time that is of the order of several $\si{ps}$ \cite{ju2004ultrafast,thiele2004spin}. In the experiments, a bias field is applied in the direction of the easy axis for a particle displaying a weak uniaxial anisotropy and the change in longitudinal magnetization $M_z$ is measured through the transient magneto-optics Kerr effect (MOKE).

To simulate the response of such particle to an ultrafast thermal pulse we consider an anisotropy parameter $\eta=0.0001$ (equivalent to an $H_K\approx0.08\si{T}$) and an applied field field of $H=0.1\si{T}$, similar to what is used in Ref.~\cite{ju2004ultrafast}. We introduce the heating produced by the thermal pulse in our model via a two temperature model (2TM) \cite{Mendil2014}, where the magnetization of the particle is coupled via the effective electron temperature $T_e$. The 2TM is defined as
\begin{subequations}
        \label{eq:2tm}
    \begin{align}
        &C_e(T)\frac{dT_e}{dt}=-G_{el}\left(T_e-T_l\right)+P(t)
        \, ,\\
        &C_l\frac{dT_l}{dt}=G_{el}\left(T_e-T_l\right)\, ,
    \end{align}
\end{subequations}
where $T_l$ is the lattice temperature, $C_e=\gamma_e T_e$ is the electron specific heat capacity and $\gamma_e$ is the electron heat capacity constant, $C_l$ is the lattice specific heat capacity, and $G_{el}$ is the electron-lattice exchange. The ultrafast laser pulse is introduced as a Gaussian pulse:
\begin{equation}
    P(t)=P_0\exp{\left[-2.77\left(\frac{t-3\tau_{pulse}}{\tau_{pulse}}\right)\right]}\, ,
\end{equation}
where $\tau_{pulse}$ is the duration of the laser pulse and $P_0$ is the nominal optical power. The parameters for the 2TM used in the simulations are given in Tab~\ref{tab:2tm_param}. The power of the pulse is chosen such that the electron temperature $T_e$ rises above the Curie temperature ($T_C=715\si{K}$) in the first $100\si{fs}$ when the pulse is applied, and $T_e$ equilibrates with $T_l$ after $\tau_{eq}=10\si{ps}$, where $T_l(\tau_{eq})$ is below the phase transition temperature $T_M\approx 350\si{K}$.
\begin{table}
\begin{tabular}{ |c|c|c|c| } 
\hline
 & Value  & Unit   \\
\hline
$\gamma_e$ & $ 3.5\times 10^{-3}$ & $\si{J.mol^{-1}.K^{-2}}$ \\ 
$C_l$ &$4.45\times 10^{1}$     & $\si{J.mol^{-1}.K^{-1}}$ \\ 
$G_{el}$ &$1.05\times 10^{12}$  & $\si{J.mol^{-1}.K^{-1}.s^{-1}}$   \\ 
$P_0$    &$1.5\times10^{16}$      & $\si{J.mol^{-1}.s^{-1}}$   \\ 
$\tau_{pulse}$    &$100$      & $\si{fs}$   \\ 
  \hline
\end{tabular}

\caption{\label{tab:2tm_param}Two temperature model parameters for \cref{eq:2tm}.}
\end{table}

\begin{figure}
    \centering
    \includegraphics[width=\textwidth]{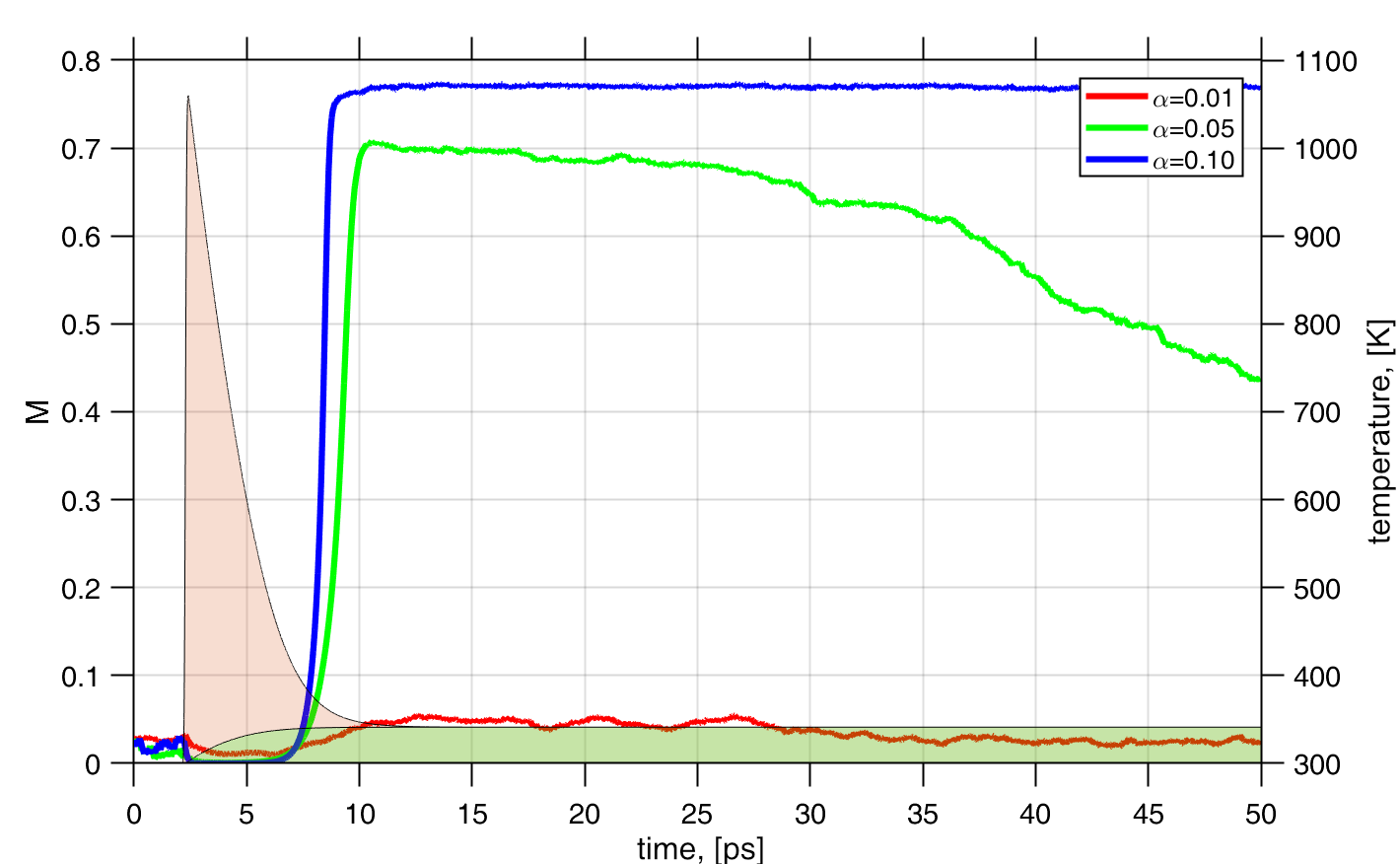}
    \caption{Time dependence of the magnetization for an isotropic particle after laser heating with a $100\si{fs}$ laser pulse for $\lambda=0.01$ (red line), $\lambda=0.01$ (green line), and $\lambda=0.1$ (blue line). The red shaded area defines the electron temperature profile and the green shaded area defines the sublattice temperature.}
    \label{fig:dynamalpha}
\end{figure}

The results for different values of the atomistic damping parameter $\lambda=0.01, 0.05,0.1$ are shown in \cref{fig:dynamalpha}. The phase transition depends on the damping parameter. In the low-damping regime ($\lambda=0.01$), the contribution of the transverse intra-sublattice exchange $\mathbf{H}_{E_\mu}^\perp$ to the perpendicular damping is not strong enough for the phase transition to occur in the time scale of the temperature pulse, which is due to the low coupling with the magnetic system. Higher damping ($\lambda=0.05$) leads to a partial phase transition into the FM phase. This FM phase transition lasts for approximately $20\si{ps}$ before decaying back to the AF phase. For still larger damping parameters ($\lambda=0.10$), the perpendicular field leads to a complete transition into the FM phase. The increased stability due to the larger equilibrium magnetization \cref{eq:mv0} after the cool down leads to the FM state to persists for hundreds of picoseconds. Increasing the damping further leads to a faster collapse into the AF phase due to the increased magnitude of the force exercised by the perpendicular intra-sublattice exchange in the perpendicular relaxation. The results obtained are consistent with what has been observed in the experimental results \cite{ju2004ultrafast} as well as the atomistic simulations \cite{barker2015higher}. 

The dynamics phase transition observed in the micromagnetic model shows a sharper transition into the FM phase for$\lambda=0.05$ than the one observed using the atomistic model. These differences can be explained by the finite dimension effects in the computation of the effective damping for small particles shown both in theory \cite{garanin1991generalized,garanin2004thermal} and numerical simulations \cite{strungaru2020model}.

The presented framework is also applicable to materials with different magnetic moments in the two sublattices, i.e., for ferrimagnetic materials. To demonstrate the model for such a case, we consider a ferrimagnetic material whith the magnetization moments in the two sublattices given by $\mu_A=3\mu_b$ and  $\mu_B=1.5\mu_b$. We also assume for the two sublattices different Curie temperature $T_{C,A}=(J_{2,A}+J_1)/3k_B$ and  $T_{C,B}=(J_{2,B}+J_1)/3k_B$. The parameters used in the simulations are given in \cref{tab:datahypothetical}.

\begin{table}
\begin{tabular}{ |c|c|c|c| } 
\hline
 & Value   & Unit   \\
\hline
$T_{C,A}$ & $ 411 $ & K \\
$T_{C,B}$ & $ 531$ & K \\ 
$J_{2,A}$ & $1.357\times10^{-20}$& J\\
$J_{2,B}$ & $1.764\times10^{-20}$& J\\
$J_{1}$ & $3.751\times10^{-22}$& J\\
$D_{\langle Q\rangle}$ &$0.808\times 10^{-22}$  & J \\
$\mu_{A}$   & $3.0$& $\mu_b$\\
$\mu_{B}$   & $1.5$& $\mu_b$\\
\hline
\end{tabular}

\caption{\label{tab:datahypothetical} material parameters for an hypothetical Heusler alloy showing a first order ferri- to ferromagnetic transition near room temperature}
\end{table}

\begin{figure}
    \centering
    \includegraphics[width=0.8\textwidth]{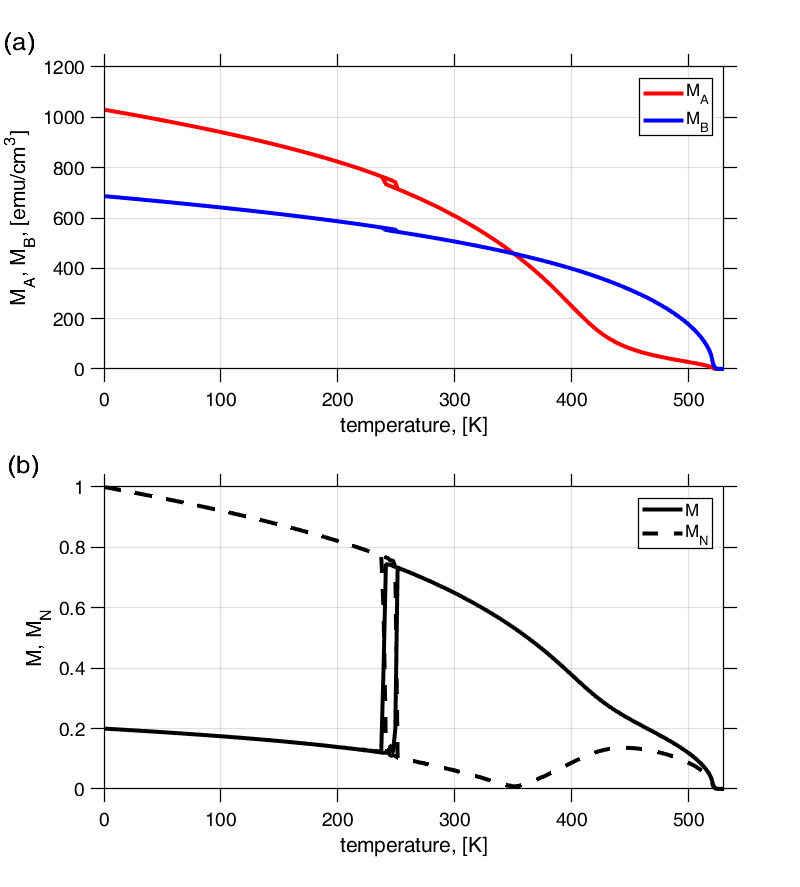}
    \caption{(a) Magnetization as a function of the temperature for the two sublattices of a ferrimagnetic material described by the parameters in \cref{tab:datahypothetical}. (b) Normalized magnetization and Néel vector for the ferrimagnetic material.}
    \label{fig:mvThypothetical}
\end{figure}

For the considered ferrimagnetic material, which has a magnetic moments $\mu_A>\mu_B$, if $T_{C,A}<T_{C,B}$, the magnetization in the the sublattice A decreases as a function of the temperature faster than the magnetization in the sublattice B, \cref{fig:mvThypothetical}(a). The material, initially in the ferrimagnetic phase, transitions to the ferrimagnetic phase,  \cref{fig:mvThypothetical}(b), when the magnetization length is below the critical magnetization for each sublattice (i.e. $m_{e,\mu}<m_{cr,\mu}$). The Curie temperature of the material is given by the largest of the Curie temperatures of the sublattices (i.e. $T_C\approx T_{C,B}$).

For a material with $J_{2,A}\neq J_{2,B}$, we cannot assume $m_{e,A}=m_{e,B}=m_e$ as we did in \cref{sec:resultenergy}, and we expect $m_{cr,A}\neq m_{cr,B}$. However, the critical magnetization $m_{cr}=\sqrt{J_1/4D{\langle Q\rangle}}$ obtained in the previous section can still be used to estimate the transition temperature of the material, since it can be observed that the critical magnetization for the faster decaying sublattice is $m_{cr,A}\lessapprox m_{cr}$. 

The Néel vector decreases until it becomes zero at the compensation temperature $T_{CP}$, where $M_{A}(T_{CP})=M_{B}(T_{CP})$. It increases for $T>T_{CP}$ when $M_B(T)>M_A(T)$ up to a maximum before going back to zero at the Curie temperature.

\section{\label{sec:simulation} Conclusions}
We presented a micromagnetic formulation for modeling ferrimagnetic materials at low and high temperatures, including cases with metamagnetic (AF to FM) phase transitions. The model is based on a mean field approximation (MFA) of the system energy that is used to derive an LLB equation. The ferrimagnet is described micromagnetically by two coupled sublattices as in the previous work by Atxitia et al. \cite{Atxitia2012}. However, our model includes one inter- and one intra-sublattice micromagnetic exchange. In addition, four-spin interactions are introduced via an inter-sublattice molecular field and a perpendicular molecular field with a cubic dependence in the magnetization of the two sublattices. The LLB equation is presented in two forms: a general form and a form simplified under the assumption of a strong homogeneous exchange field, which is applicable to most ferromagnetic and ferrimagnetic materials.

The presented formulation was used for modeling the thermal equilibrium and metamagnetic phase transitions in FeRh. 
The simulations show that the origin of such transitions is in the inter-sublattice molecular field obtained from the nearest-neighbors and second-nearest neighbors as well as the molecular field with cubic dependence in the magnetization obtained from the four-spin interactions \cite{barker2015higher,mryasov2005}. The formulation reproduces the hysteretic AF to FM transition behaviour and time scales observed in recent experiments \cite{thiele2003ferh,kouvel1962anomalous,ju2004ultrafast} and atomistic simulations \cite{barker2015higher}. 

The model we developed can be considered as an extension of previous micromagnetic models and it is able to simulate ferrimagnetic materials showing similar first-order phase transitions, like Heusler alloys \cite{ovichi2014ferri}, and it can be used to model a wide range of ferrimagnetic materials and phenomena, including recently observed all-optical driven THz spintronic effects observed in FeRh \cite{medapalli2020femtosecond,menarini2019micromagnetic} as well as memory application that exploit phase transitions \cite{fina2020flexible}. 
\section{\label{sec:aknowledgment} Acknowledgments}
This work was supported as part of the Quantum-Materials for Energy Efficient Neuromorphic-Computing(Q-MEEN-C), an Energy Frontier Research Center funded by the U.S. Department of Energy, Office of Science, Basic Energy Sciences under Award No. DE-SC0019273. The authors thank Professor Prof Roy Chantrell and Dr. Mara Strungaru for the help with the atomistic modelling as well as Dr. Joseph Barker for the helpful conversation. For simulations, this work used the Extreme Science and Engineering Discovery Environment (XSEDE), which is supported by National Science Foundation grant number ACI-1548562, specifically, it used the Bridges and Comet systems supported by NSF Grant \# ACI-1445506 at Pittsburgh and San Diego Supercomputer Centers.
\newpage
\medskip

\providecommand{\noopsort}[1]{}\providecommand{\singleletter}[1]{#1}%

%\printbibliography

\end{document}